\documentclass[a4paper,twocolumn,english,conference]{IEEEtran}
\usepackage[T1]{fontenc}
\usepackage[latin9]{inputenc}
\usepackage{verbatim}
\usepackage{float}
\usepackage{bm}
\usepackage{amsmath}
\usepackage{amsthm}
\usepackage{amssymb}
\usepackage{graphicx}
\usepackage{wasysym}

\makeatletter

\pdfpageheight\paperheight
\pdfpagewidth\paperwidth

\floatstyle{ruled}
\newfloat{algorithm}{tbp}{loa}
\providecommand{\algorithmname}{Algorithm}
\floatname{algorithm}{\protect\algorithmname}

\theoremstyle{plain}
\newtheorem{thm}{\protect\theoremname}
\theoremstyle{definition}
\newtheorem{defn}[thm]{\protect\definitionname}

\usepackage[english]{babel}
\usepackage[noadjust]{cite}
\usepackage{times}
\usepackage[usenames,dvipsnames]{xcolor}
\usepackage{amsfonts}
\usepackage{psfrag}
\usepackage{fancyhdr}
 \usepackage{algorithmic}
\allowdisplaybreaks

\usepackage{soul}

\providecommand{\definitionname}{Definition}
\providecommand{\theoremname}{Theorem}

\usepackage{babel}
\providecommand{\definitionname}{Definition}
\providecommand{\theoremname}{Theorem}

\usepackage{babel}
\providecommand{\definitionname}{Definition}
\providecommand{\theoremname}{Theorem}

\makeatother

\usepackage{babel}
\providecommand{\definitionname}{Definition}
\providecommand{\theoremname}{Theorem}

\begin{document}
\bstctlcite{IEEEWCNC:BSTcontrol}

\title{\vspace{-1.25cm}
 Edge Computing Meets Millimeter-wave Enabled VR: Paving the Way to
Cutting the Cord\vspace{-0.5cm}
 }

\author{\IEEEauthorblockN{Mohammed~S.~Elbamby\IEEEauthorrefmark{1}, Cristina Perfecto\IEEEauthorrefmark{2},
Mehdi~Bennis\IEEEauthorrefmark{1}, and Klaus Doppler\IEEEauthorrefmark{3}\vspace{-0.05cm}
 } \IEEEauthorblockA{\IEEEauthorrefmark{1}Centre for Wireless Communications, University
of Oulu, Finland. \vspace{-0.05cm}
 \\
 emails: \{mohammed.elbamby,mehdi.bennis\}@oulu.fi \vspace{-0.05cm}
 \\
 }\IEEEauthorblockA{\IEEEauthorrefmark{2}University of the Basque Country (UPV/EHU),
Spain. email: cristina.perfecto@ehu.eus \vspace{-0.05cm}
 \\
 }\IEEEauthorblockA{\IEEEauthorrefmark{3}Nokia Bell Labs, Sunnyvale, CA, USA. email:
klaus.doppler@nokia-bell-labs.com\vspace{-0.15cm}
 \\
 }}
\maketitle
\begin{abstract}
In this paper, a novel proactive computing and mmWave communication
for ultra-reliable and low latency wireless virtual reality (VR is
proposed. By leveraging information about users' poses, proactive
computing and caching are used to pre-compute and store users' HD
video frames to minimize the computing latency. Furthermore, multi-connectivity
is exploited to ensure reliable mmWave links to deliver users' requested
HD frames. The performance of the proposed approach is validated on
a VR network serving an interactive gaming arcade, where dynamic and
real-time rendering of HD video frames is needed and impulse actions
of different players impact the content to be shown. Simulation results
show significant gains of up to $30\%$ reduction in end-to-end delay
and $50\%$ in the $90^{\textrm{th}}$ percentile communication delay. 
\end{abstract}


\vspace{-0.2cm}

\section{Introduction}

\vspace{-0.1cm}


Commercial 5G deployments are not expected to be available before
2020, however the race to showcase a first pre-standard 5G network
is on. Hence, all eyes are set on how brand new services that promise
to deliver entirely new experiences such as 360-degree immersive virtual
reality (VR) will be offered. However multiple technical challenges
need to be investigated to deal with the latency-sensitivity and the
resource \textendash communications and computing\textendash{} intensiveness
nature of 4K/8K UHD immersive VR wireless streaming and to realize
the vision of interconnected VR \cite{ejder_VR_2017}.

To accommodate the extensive use of resource-hungry applications,
it is foreseen that a 1000-fold boost will be needed in system capacity
(measured in bps/km$^{2}$). This will be facilitated via an increased
bandwidth, higher densification, and improved spectral efficiency.
Zooming on the VR requirements, even anticipating the use of 265 HEVC
1:600 video compression rate, a bit rate of up to 1~Gbps \cite{VR_NET_mag}
would be needed to match the 2x64 million pixel human-eye accuracy.
These rates are unrealizable in 4G and challenging in 5G for a disruption-free
immersive VR demands. Therefore significant research efforts around
VR have focused on reducing bandwidth needs in mobile/wireless VR,
thereby shrinking the amount of data processed and transmitted. Many
approaches leverage head and eye-gaze tracking to spatially segment
360$^{\circ}$ frames and deliver in HD only user's field of view
(FOV) matching portion \cite{qian_optimCell_2016}, \cite{ju_ultraWideVRStream_2017}.
Alternatively \cite{doppler_EUCNC_2017} considers a foveated 360$^{\circ}$
transmission where resolution and color depth are gradually reduced
from fovea-centralis area to the peripheral FOV.

Latency is critical for VR; the human eye needs to experience accurate
and smooth movements with low (<20~ms) motion-to-photon (MTP) latency
\cite{ju_ultraWideVRStream_2017,doppler_EUCNC_2017,Walid_VR}. High
MTP values send conflicting signals to the Vestibulo-ocular reflex
(VOR), and might lead to dizziness or motion sickness. Both computing
(image processing or frame rendering) and communication (queuing and
over-the-air transmission) delays represent a major bottleneck in
VR systems. Heavy image processing requires high computational power
that is often not available in the local head-mounted device (HMD)
GPUs. Offloading computing significantly relieves the computing burden
at the expense of incurring an additional communication delay in the
downlink (DL) delivery of the processed video frames in full resolution.
Moreover, to ensure responsiveness and real-time computing through
minimal latency, computing servers should be readily available and
located close to the end users.

VR demands a perceptible image-quality degradation-free uniform experience.
However, temporary outages due to impairments in measured signal to
interference plus noise ratio (SINR) are frequent in mobile/wireless
environments. In this regard, an ultra-reliable VR service refers
to the timely delivery of video frames with high success rate. Provisioning
for a higher reliability pays a toll on the use of resources and allocating
more resources for a single user could potentially impact the experienced
latency of the remaining users.

Leveraging ubiquitous caching and computing at the wireless network
edge will radically change the future mobile network architecture
\cite{BigDataCaching2016}, and alleviate the current bottleneck for
massive content delivery. Research ideas and network engineering geared
toward exploiting communications, caching, and computing (C$^{3}$)
for future content-centric mobile networks are found in \cite{bonomi_fog_its_role_2012,elbamby_proactivefog_2017,mao_MEC_survey_2017}.
This paper exploits the C$^{3}$ paradigm to provide an enhanced wireless
VR experience. To that end, it blends together the use of millimeter-wave
(mmWave) technology and fog computing. The former seeks to deliver
multi-Gbps wireless communication between VR headsets and network
access points, with reliability guarantees, and the latter carries
out advanced image processing, effectively offloading client displays
or game consoles while satisfying stringent latency constraints. The
main contribution of this paper is to propose a joint proactive computing
and mmWave resource allocation scheme under latency and reliability
constraints. Reliability is ensured by leveraging multi-connectivity
(MC) to enhance the performance of users under channel variability,
whereas proactive computing and user association is optimized to satisfy
the latency requirements.

The rest of this paper is organized as follows. Section~\ref{sec:SysMod}
describes the system model and problem formulation. The proposed joint
computing and matching scheme is introduced in Section~\ref{sec:JointTandM}.
Section~\ref{sec:SimRes} analyzes the performance of the proposed
framework. Finally, Section~\ref{sec:Conc} concludes the paper.


\section{System Model\label{sec:SysMod}}

\vspace{-0.1cm}

\begin{figure}[t]
\centering\includegraphics[width=0.9\columnwidth]{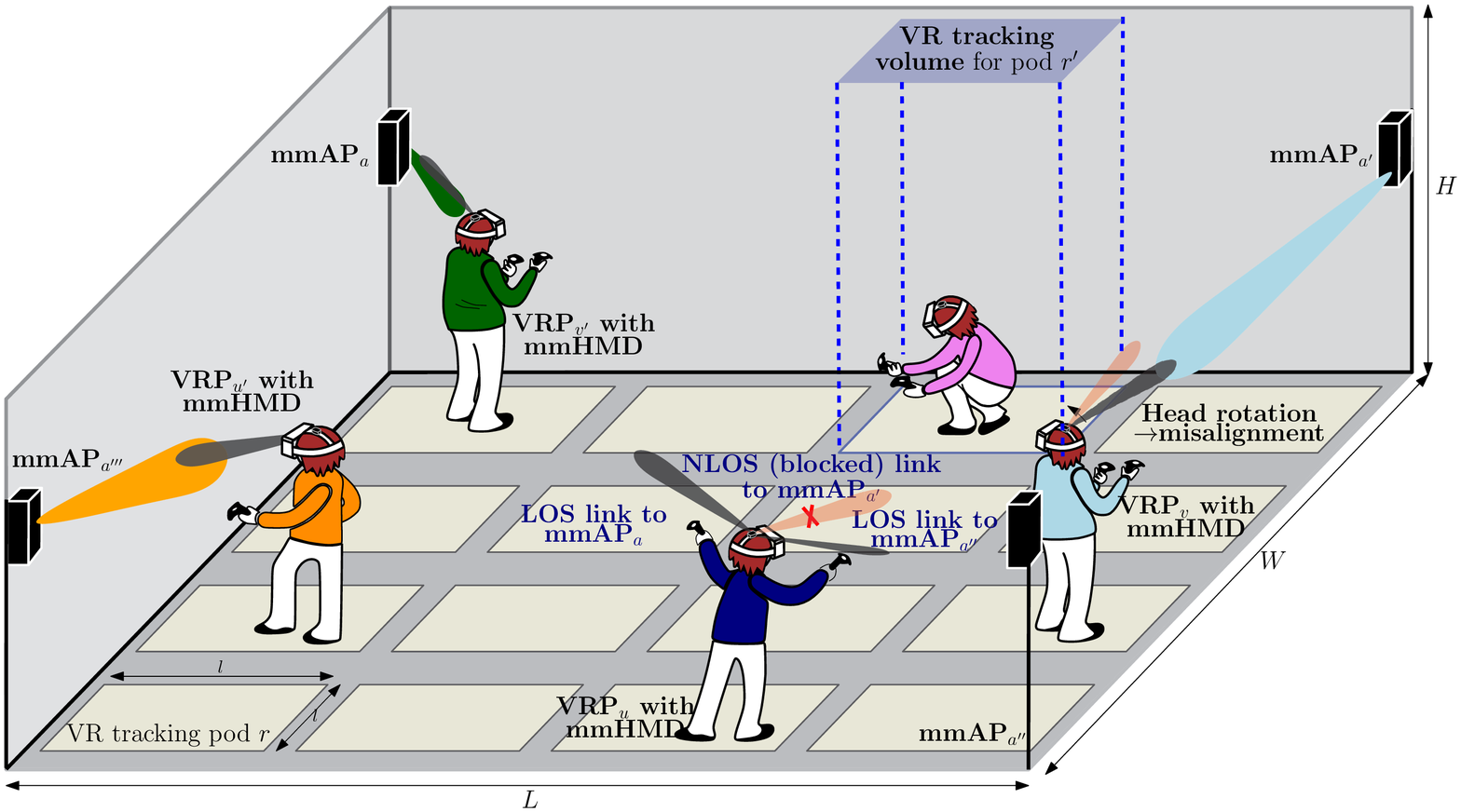}
\caption{Open-plant VR gaming arcade with VRPs moving freely within their VR
tracking pods and getting HD video frames from mmAP. }
\label{fig:sys_model} 
\end{figure}

Consider an indoor, open plan, VR gaming arcade of dimensions $L$~x~$W$~x~$H$~m$^{3}$
where a set \textbf{$\mathcal{A}$} of $A$ mmWave band access points
(mmAP) serves a set \textbf{$\mathcal{U}$} of $U$ virtual reality
players (VRP) equipped with wireless mmWave head-mounted VR displays
(mmHMD). VRPs are distributed and move freely within the limits of
the $R$ individual and $l$ x $l$ m$^{2}$ sized VR pods such that
$R\geq U$. The movement of VRPs in the physical space of each VR
pod is tracked and mapped into the virtual space. An illustration
of the system model is shown in Fig.~\ref{fig:sys_model}. The system
operates in a time-slotted mode. The time-slots are indexed by $t\in\{1,2,\cdots\}$
with separate scales to account for scheduling decisions, and for
beam alignment.

\vspace{-0.2cm}

\subsection{Interactive VR frame rendering model}

\vspace{-0.1cm}

A set \textbf{$\mathcal{I}$} of impulse actions is defined as the
actions generated during the interactive gaming either by the VRPs
or an external trigger. The arrival of 
impulse action $m_{i}\!\in\mathcal{I}$ impacts the game play of a
subset of VRPs $\mathcal{U}_{i}\subseteq\mathcal{U}$. In this regard,
the impact of the impulse actions on the VRPs' game play, namely,
the \emph{impact matrix,} is defined as $\Theta=[\theta_{ui}]$, where
$\theta_{ui}=1$ if $u\in\mathcal{U}_{i}$, and $\theta_{ui}=0$ otherwise.\footnote{An example of an impulse action is a player firing a gun in a shooting
game. As the game play of a subset of players is affected by this
action, a computed video frame for any of them needs to be rendered
again.}


Mobile edge computing is used to provide the required computation
capabilities. 
Accordingly, a fog network consisting of a set $\mathcal{E}$ of $E$
edge servers with GPU computing capability $c_{e}$, and a storage
unit of capacity $S$ HD frames, will perform real-time VR environment
building related computing\footnote{The terms computing and rendering are used interchangeably throughout
the paper to describe the process of rendering HD video frames. } to generate HD interactive video frames based on the real-time 6D
pose\footnote{A 6D pose is jointly given by the 3D location coordinates and the
3 orientation angles over the X, Y, and Z axes, namely the roll, pitch
and yaw.} and the impulse actions of VRPs.

\begin{figure*}[t]
\centering \includegraphics[width=0.7\textwidth]{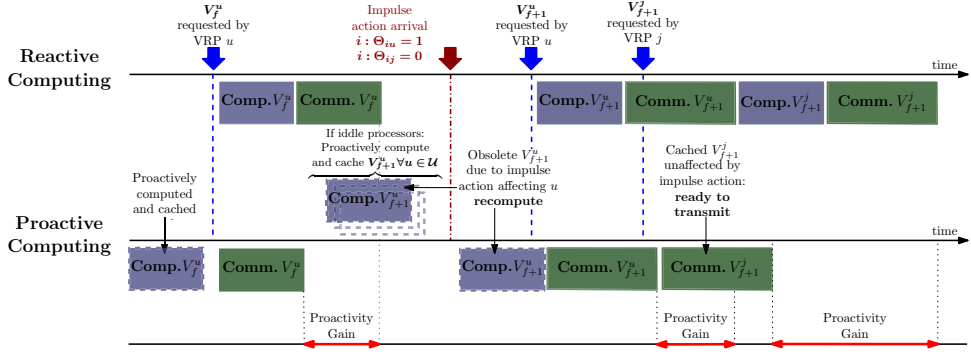}
\caption{Time dynamics for reactive and proactive computing.}
\label{fig:proactive_comp} 
\end{figure*}

Each VRP $u$ is interested in receiving a unique $F$-tuple of HD
video frames $(V_{1}^{u},V_{2}^{u}\ldots,V_{F}^{u})$ throughout its
game play. To render an HD video frame, a processing density of $\kappa$
GPU cycles per bit of frame data is required, with HD frame $V_{f}^{u}$
having a data size of $L_{fu}^{\textrm{HD}}$ bits. In this context,
mmAPs act as a two-way middleware between the mmHMDs and the edge
servers in the fog network. mmAPs relay pose and action inputs from
VRPs arriving through the wireless uplink (UL). After the HD frames
are rendered, the mmAPs schedule mmWave time slots in the DL to deliver
the resulting video frames. To ensure reliable DL transmission, MC
is leveraged in which multiple mmAPs can jointly transmit the same
data to a player with a weak link. Furthermore, to avoid motion sickness
associated with high MTP delay, computing and scheduling decisions
have to guarantee stringent latency constraints. Therefore, The fog
network leverages the predictability of users' poses to proactively
compute the upcoming HD video frames within a prediction window $T_{\textrm{w}}$.
VRPs can receive and stream the proactively computed video frames
as long as they are not affected by impulse actions arriving afterwards.
The proactive computing time dynamics are depicted in Fig. \ref{fig:proactive_comp}.
{} Moreover, edge servers exploit their idle times to proactively render
and cache the HD frames of users affected by the popular impulse actions,
such that the computing latency is minimized. Finally, it is assumed
that a low quality (LQ) version of the video frame, with data size
$L_{fu}^{\textrm{LQ}}\ll L_{fu}^{\textrm{HD}}$, can be processed
locally in the mmHMD to ensure smooth game play if the HD video frame
cannot be delivered on time. %

\vspace{-0.2cm}

\subsection{mmWave communication model}

\vspace{-0.1cm}

The mmWave channel is based on measurement results of LOS and NLOS
paths for the 60-GHz indoor channels \cite{channelRappaport}, and
includes both pathloss attenuation $l_{au}$ and small Nakagami fading
with coefficient $g_{au}(t)$. 
The channel gain $h_{au}$ from mmAP $a$ to mmHMD in VRP $p$ is
thus given by $|h_{au}(t)|^{2}=l_{au}|g_{au}(t)|^{2}$.

The pathloss $\ell_{au}$ is considered LOS when $\nexists$ VRP $u'\in\mathcal{U}\backslash u$
such that the area defined by the $d$-diameter circle associated
to its head+mmHMD intercepts the ray traced from mmAP $a$ to mmHMD
receiver of VRP $u$; the path is considered NLOS otherwise\footnote{Capturing the blockage arriving from game-engagement related limb
movements, e.g. VRPs raising their hands and blocking their own or
some other mmWave link is left for future work.}. Neither external interference leakage into the finite area of the
arcade nor reflections due to walls, ceiling or the VRPs themselves
are explicitly incorporated in the channel model. They are only accounted
for in a coarse way through the different LOS and NLOS parameters.
Accordingly, the pathloss exponent $\alpha_{au}$ in $l_{au}$ will
take value $\alpha_{L}$ if the link from mmAP $a$ to the mmHMD receiver
$u$ is LOS and value $\alpha_{N}$ otherwise. Similarly, the corresponding
Nakagami shape factor $m_{au}$ will take value $m_{L}$ for LOS and
$m_{N}$ for NLOS paths; it is further assumed that $g_{au}(t)$ is
i.i.d. and not temporally correlated.

For tractability, the radiation pattern of actual directional antennas
is approximated with a 2D sectored antenna model \cite{wildman_2DsecAntenna_2014}.
Let $g_{au}^{\textup{Tx}}(\varphi_{a},\vartheta_{up}^{\textup{Tx}}(t))$
and $g_{au}^{\textup{Rx}}(\varphi_{u},\vartheta_{au}^{\textup{Rx}}(t))$
denote the transmission and reception antenna gains from mmAP $a$
to the mmHMD of VRP $u$ as given by 

\vspace{-0.4cm}
 
\begin{equation}
g_{au}^{\varangle}(\varphi_{a},\vartheta_{au}^{\varangle}(t))=\left\lbrace \begin{array}{ll}
\frac{2\pi-\left(2\pi-\varphi_{a}\right)g_{sl}}{\varphi_{a}}\text{,} & |\vartheta_{au}^{\varangle}(t)|\leq\frac{\varphi_{a}}{2},\\
\vspace*{-0.1cm}g_{sl}, & \text{otherwise,}
\end{array}\right.\label{eq:antennaTx}
\end{equation}

\vspace{-0.2cm}
 \setlength{\parindent}{0pt}with $\varangle\in\{\textup{Tx},\textup{Rx}\}$,
and where $\vartheta_{au}^{\varangle}(t)$ stands for the angular
deviation from the boresight directions, 
and $g_{sl}$ is the constant sidelobe gain with $g_{sl}\ll1.$ 


High directionality of mmWave communications motivates a search process
to find the boresight directions corresponding to the best path between
the mmAP and the mmHMD and take full advantage of beamforming gains
as per \eqref{eq:antennaTx}. 
To that end, periodically each of the $A$ mmAPs will sequentially
perform beam training with the $|\mathcal{U}_{a}|$ VRPs within it
coverage area, $\mathcal{U}_{a}\subset\mathcal{U}$. After the best
steerings for all feasible VRPs have been learned, and following a
transmission interval level VRP scheduling decision, data transmission
begins.

The analog beamformers on the mmAPs and VRPs sides are determined
after a two-stage beam training process \cite{wang_beamtraining_2009}.
Let $\bm{\tau}=\{\tau_{1},\tau_{2},\cdots,\tau_{A}\}$ denote the
vector of alignment delays for the $A$ mmAPs in the system, then
on an \emph{a priori} knowledge of mmAPs fixed location and of VRPs
sector\footnote{A reasonable assumption due to highly accurate and frequent VRP location
tracking and to the slowness of human movement at $ms$ scale.}, 
the experienced alignment delay $\tau_{au}$ due to beam training
is $\tau_{au}=\frac{\psi_{a}}{\varphi_{a}}T_{p}$, where $T_{p}$
is the pilot symbol transmission time and $\psi_{a}$, $\varphi_{a}$
denote the sector-level and beam-level beamwidths for mmAP $a$. The
overall alignment delay is $\tau=\sum_{\substack{a\in\mathcal{A}}
}\sum_{\substack{u\in\mathcal{U}_{a}}
}\tau_{au}$.

To overcome vulnerability to channel intermittency due to blockage
and misalignment, mmHMDs are assumed to be capable of leveraging MC
implemented through coordinated joint transmission of VR player data
from multiple mmAPs. An indicator variable $x_{au}(t)$ is therefore
defined if mmAP $a$ schedules VRP $u$ at time instant $t$. The
maximum achievable rate (in Gbps) for mmHMD $u$ is given by \vspace{-0.2cm}

\begin{equation}
r_{u}(t)=\left(1-\tau\right)B\log_{2}\left(1+\gamma_{u}(t)\right),\label{eq:rate}
\end{equation}
\begin{equation}
\negthinspace\gamma_{u}(t)\negthinspace=\negthinspace\frac{\sum_{\substack{\forall a\in\mathcal{A}}
}x_{au}(t)p_{a}|h_{au}(t)|^{2}g_{au}^{\textup{Tx}}(t)g_{au}^{\textup{Rx}}(t)}{\negthinspace\sum_{\substack{\forall a'\in\mathcal{A}}
}(1\negthinspace-\negthinspace x_{au}\negthinspace(t))p_{a'u}|h_{a'u}\negthinspace(t)|^{2}g_{a'u}^{\textup{Tx}}\negthinspace(t)g_{a'u}^{\textup{Rx}}\negthinspace(t)\negthinspace+\negthinspace N_{0}B},
\end{equation}
where the achievable SINR term $\gamma_{u}(t)$ should, in addition
to the effective received power at mmHMD $u$ from mmAPs $a:x_{au}(t)=1$
and to Gaussian noise, account for the effect of other interfering
mmAPs $a':x_{a'u}(t)=0$ through channel and antenna gains, $|h_{a'u}(t)|^{2}$
and $g_{a'u}^{\textup{Tx}}$, $g_{a'u}^{\textup{Rx}}$ respectively.

\vspace{-0.2cm}

\subsection{Computing model}


The total perceived delay to compute and deliver an HD video frame
is expressed as: \vspace{-0.2cm}

\begin{equation}
D_{uf}(t)=\xi_{fu}(D_{uf}^{\textrm{cp}}(t)+D_{uf}^{\textrm{cm}}(t)+\text{\ensuremath{\tau}}_{\textrm{EP}}),
\end{equation}

\vspace{-0.2cm}
 where $\xi_{fu}$ is a binary function that equals $1$ when the
HD video frame is delivered to VRP $u$ and equals $0$ if the LQ
frame is delivered, $D_{uf}^{\textrm{cp}}$ and $D_{uf}^{\textrm{cm}}$
are the computing and communication delays of HD frame $f$ initiated
from user $u$, and $\text{\ensuremath{\tau}}_{\textrm{EP}}$ is the
processing latency which accounts for the edge server processing,
storage processing and the UL transmission of user pose and action
data. Here, we focus on the effect of computation delay in the edge
servers and the DL communication delay, i.e., the access to the backhaul
is assumed to be wired. Let the computing delay $D_{uf}^{\textrm{cp}}$
be expressed as follows: \vspace{-0.1cm}
 
\begin{equation}
D_{uf}^{\textrm{cp}}(t)=\biggl(\frac{\kappa L_{fu}^{\textrm{HD}}}{c_{e}}+W_{uf}(t)\biggr)z_{fu}(t)(1-y_{fu}(t)),
\end{equation}

\vspace{-0.2cm}
 where $c_{e}$ is the computation capability of edge server $e$,
$z_{fu}(t)$ and $y_{fu}(t)$ indicate that the video frame $f$ of
user $u$ is scheduled for computing, and is cached in the fog network
at time instant $t$, respectively, and $W_{uf}$ is the computation
waiting time of HD frame $f$ of user $u$ in the service queue, defined
as $Q(t)$. Furthermore, Let the communications delay $D_{uf}^{\textrm{cm}}$
be as follows:

\vspace{-0.4cm}
 
\begin{equation}
D_{uf}^{\textrm{cm}}(t)\hspace{-0.7mm}=\hspace{-0.7mm}\arg\min_{d_{u}}\hspace{-2mm}\sum_{t'=D_{uf}^{\textrm{cp}}(t)+1}^{D_{uf}^{\textrm{cp}}(t)+d_{u}}\hspace{-0.5mm}\biggl(T_{t}r_{u}(t')\geq L_{fu}^{\textrm{HD}}\biggr),
\end{equation}

\vspace{-0.2cm}
 where the \emph{$\arg\min$} function is to find the minimum number
of time slots needed for the video frame $f$ to be delivered.

The objective of the proposed approach is to maximize the HD video
frame delivery $\boldsymbol{\xi}=[\xi_{fu}]$ subject to latency constraints.
The optimization variables are the scheduling, caching, and computing
matrices, expressed as $\boldsymbol{X}(t)=[x_{ua}(t)]$, $\boldsymbol{Y}(t)=[y_{fu}(t)]$,
and $\boldsymbol{Z}(t)=[z_{fu}(t)]$ respectively. The optimization
problem is cast as follows: \vspace{-2mm}
 \begin{subequations}\label{eq:objective} 
\begin{align}
\max_{\boldsymbol{X}(t),\boldsymbol{Y}(t),\boldsymbol{Z}(t)} & \sum_{u=1}^{U}\sum_{f=1}^{F}\xi_{fu}\\
\parbox{1.6cm}{\text{subject to}} & \Pr(D_{uf}(t)\hspace{-0.5mm}\geq\hspace{-0.5mm}D_{\textrm{th}})\hspace{-0.8mm}\leq\hspace{-0.5mm}\epsilon,\;\hspace{-0.8mm}\forall\hspace{-0.5mm}f\hspace{-0.5mm}\in\hspace{-0.5mm}\mathcal{F},\forall u\hspace{-0.5mm}\in\hspace{-0.5mm}\mathcal{U},\label{eq:prob_const}\\
 & \sum_{u=1}^{U}x_{ua}(t)\leq1,\;\forall a\in\mathcal{A},\label{eq:match_constraint_2}\\
 & \sum_{u=1}^{U}\sum_{f=1}^{F}y_{fu}(t)\leq S\;,\label{eq:cache_const}\\
 & \sum_{u=1}^{U}\sum_{f=1}^{F}z_{fu}(t)\leq E,\label{eq:comp_const}
\end{align}
\end{subequations} \vspace{-0.1cm}
 where (\ref{eq:prob_const}) is a probabilistic delay constraint
that ensures the communication latency is bounded by a threshold value
$D_{\textrm{th}}$ with a probability $1-\epsilon$. Constraint (\ref{eq:match_constraint_2})
ensures that mmAP serves one VRP at a time. (\ref{eq:cache_const})
limits the number of cached frames to a maximum of $S$. Constraint
(\ref{eq:comp_const}) is over the maximum number of simultaneous
computing processes. The above problem is a combinatorial problem
with non-convex cost function and probabilistic constraints, for which
finding an optimal solution is computationally complex \cite{QoS_const}.
The non-convexity is due to the interference from other mmAPs in the
rate term in the delay equation. To make the problem tractable, we
use the Markov's inequality to convert the probabilistic constraint
in (\ref{eq:prob_const}) to a linear constraint \cite{QoS_const}
expressed as $\mathbb{E}\{D_{uf}(t)\}\leq D_{\textrm{th}}\epsilon$.
Hence, it can be rewritten as: \vspace{-0.6cm}

\begin{multline}
\negthinspace\mathbb{E}\negthinspace\left\lbrace \negthinspace D_{uf}^{\textrm{cp}}(t)\negthinspace+\negthinspace D_{uf}^{\textrm{cm}}(t)\negthinspace\right\rbrace \negthinspace=\negthinspace\mathbb{E}\negthinspace\left\lbrace \negthinspace D_{uf}^{\textrm{cp}}(t)\negthinspace\right\rbrace \negthinspace+\negthinspace\mathbb{E}\negthinspace\left\lbrace \negthinspace D_{uf}^{\textrm{cm}}(t)\negthinspace\right\rbrace \negthinspace\leq\negthinspace D_{\textrm{th}}\epsilon
\end{multline}

\vspace{-0.1cm}
 The above expectation is hard to calculate due to having the $\arg\min$
function in the $D_{uf}^{\textrm{cm}}(t)$. Therefore, we express
the second expectation as $\mathbb{E}\left\lbrace D_{uf}^{\textrm{cm}}(t)\right\rbrace =\frac{L_{fu}^{\textrm{HD}}}{T_{t}\bar{r}_{u}(t)},$
where $\bar{r}_{u}(t)$ is the time-average service rate of VRP $u$
from mmAP $p$ at time instant $t$, expressed as $\bar{r}_{u}(t)=\sum_{\tau=1}^{t}r_{a}(\tau)$.
Hence, the ultra-reliable low latency communication (URLLC) constraint
can be rewritten as: 
\begin{equation}
\begin{split}\frac{L_{fu}^{\textrm{HD}}}{T_{t}\bar{r}_{u}(t)}+\frac{\kappa L_{fu}^{\textrm{HD}}}{c_{e}}+\mathbb{E}\left\lbrace W_{uf}(t)\right\rbrace \leq D_{\textrm{th}}\epsilon.\end{split}
\label{eq:const_makov}
\end{equation}
Similarly, the average waiting time can be expressed as $\mathbb{E}\left\lbrace W_{uf}(t)\right\rbrace =\sum_{i|V_{f}^{i}\in Q(t)}\frac{L_{fi}^{\textrm{HD}}}{T_{t}\bar{r}_{i}(t)}$,
and (\ref{eq:const_makov}) rewritten as:\vspace{-0.2cm}
 
\begin{equation}
\frac{L_{fu}^{\textrm{HD}}}{T_{t}\bar{r}_{u}(t)}\leq D_{\textrm{th}}\epsilon-\hspace{-2mm}\sum_{V_{f}^{i}\in Q(t)}\frac{L_{fi}^{\textrm{HD}}}{T_{t}\bar{r}_{i}(t)}-\frac{\kappa L_{fu}^{\textrm{HD}}}{c_{e}}.\label{eq:final_delay_const}
\end{equation}

\vspace{-0.2cm}

\begin{algorithm}[t]
\begin{algorithmic}[1]\footnotesize

\STATE  \textbf{\footnotesize{}Implementation at each time instant
$t$:}{\footnotesize{} }{\footnotesize \par}

\STATE  \textbf{\footnotesize{}Repeat}{\footnotesize{} find v = vacant
cloudlet}{\footnotesize \par}

\hspace{-0.6cm} \textbf{\emph{\footnotesize{}Priority one: real-time
scheduling}}{\footnotesize \par}

\STATE  {\footnotesize{}find $u$ where $V_{f}^{u}(t)$ is not computed}{\footnotesize \par}

\STATE  {\footnotesize{}$v\rightarrow allocated$, $z_{fu}(t)\rightarrow1$}{\footnotesize \par}

\hspace{-0.6cm} \textbf{\emph{\footnotesize{}Priority two: predictive
computing and caching}}{\footnotesize \par}

\STATE  {\footnotesize{}Find $u$ and $t'$ where $t'\in(t:t+T_{\textrm{w}})$
and $V_{f}^{u}(t')$ is not computed}{\footnotesize \par}

\STATE  {\footnotesize{}$v\rightarrow allocated$, $z_{fu}(t)\rightarrow1$}{\footnotesize \par}

\STATE  {\footnotesize{}compute and cache $V_{f}^{u}(t')$ }{\footnotesize \par}

\hspace{-0.6cm} \textbf{\emph{\footnotesize{}Priority three: predictive
impulse computing and caching}}\textbf{\footnotesize{} }{\footnotesize \par}

\STATE  {\footnotesize{}Sort $\mathcal{I}$ by popularity}{\footnotesize \par}

\STATE  \textbf{\footnotesize{}Repeat}{\footnotesize{} Select $i$
as the most popular $\mathcal{I}$}{\footnotesize \par}

\STATE  {\footnotesize{}find $u$:$\theta_{u,i}=1$}{\footnotesize \par}

\STATE {\footnotesize{}$v\rightarrow allocated$, $z_{fu}(t)\rightarrow1$}{\footnotesize \par}

\STATE {\footnotesize{}compute and cache $V_{f}^{u}(t+1)$}{\footnotesize \par}

\STATE \textbf{\footnotesize{}Until}{\footnotesize{} at least one
of the following conditions is true }{\footnotesize \par}
\begin{itemize}
\item {\footnotesize{}no vacant cloudlets }{\footnotesize \par}
\item {\footnotesize{}cache is full }{\footnotesize \par}
\item {\footnotesize{}upcoming frames of impacted users from all $\mathcal{I}$
are computed }{\footnotesize \par}
\end{itemize}
\end{algorithmic}

\caption{\label{alg:CandC_algo}Joint computing and caching algorithm.}
\end{algorithm}

The average rate $\bar{r}_{u}(t)$ can be separated into the instantaneous
time rate at time instant $t$ and the average rate in the previous
time instants (that can be estimated), i.e., $\bar{r}_{u}(t)=\sum_{\varrho=1}^{t}r_{u}(\varrho)=r_{u}(t)+\sum_{\varrho=1}^{t-1}r_{u}(\varrho)$.
In other words, to reach the desired latency requirement, a maximum
value of $\frac{L_{fu}^{\textrm{HD}}}{T_{t}\bar{r}_{u}(t)}$ that
satisfies the above formula should be guaranteed to admit admitted
a request to an mmAP.

Next, a joint VRP-mmAP matching and scheduling scheme is proposed
to solve the optimization problem in (\ref{eq:objective}).


\section{Joint Proactive Computing and Matching\label{sec:JointTandM} }



As stated above, the optimization problem in (\ref{eq:objective})
is computationally hard to solve. We rather decouple it into two subproblems
of computation scheduling and VRP association.

\subsection{Computing and caching scheme}


During the game play, the edge computing network minimizes the computing
service delay by leveraging the pose prediction of users to proactively
compute their HD video frames as well as caching the updated HD video
frames resulting from randomly arriving impulse actions. In the case
that impulse action arrives in which the corresponding frames for
the affected users are not computed, the real time computing is giving
the highest priority. Therefore, we propose a three priority level
algorithm to schedule HD frame computing of users. The detailed algorithm
is described in Algorithm \ref{alg:CandC_algo}.


\subsection{Player-Server matching }

Our next step is to propose a VRP-mmAP association scheme that solves
the constrained minimization problem in (\ref{eq:objective}). 
The association problem is formulated as a matching game \cite{OmidMatching2014}
between the mmAPs and the VRPs. In this game, VRPs seek to maximize
their VR experience by competing for mmWave time slots from different
mmAPs. Whenever a player in the network requests a new HD frame, a
new set of matching pairs is found using the proposed approach. The
matching game consists of a two sets of players and mmAPs, where each
member of one set has a preference profile over the members of the
other sets. Preferences of mmAPs and players are denoted by $\succ_{a}$
and $\succ_{u}$, and reflect how each member of a set ranks the members
of the other set.

\begin{algorithm}[t]
\begin{algorithmic}[1]\footnotesize

\STATE {\footnotesize{} }\textbf{\footnotesize{}Initialization: }{\footnotesize{}all
players and mmAPs start unmatched.}{\footnotesize \par}

\STATE {\footnotesize{} Each mmAP constructs its preference list
as per (\ref{eq:cloudlet_preference})}{\footnotesize \par}

\STATE  {\footnotesize{}Each player constructs its preference list
as per (\ref{eq:user_preference})}{\footnotesize \par}

\STATE {\footnotesize{} }\textbf{\footnotesize{}repeat}{\footnotesize{}
an unmatched player $u$, i.e., $\Upsilon(u)=\phi$ proposes to its
most preferred mmAP $a$ that satisfies $a\succ_{u}u$}{\footnotesize \par}

\begin{ALC@g}\STATE {\footnotesize{} }\textbf{\footnotesize{}if}{\footnotesize{}
$\Upsilon(a)=\phi$, ~\%not yet matched}{\footnotesize \par}

\begin{ALC@g}\STATE {\footnotesize{}$\Upsilon(a)=u$, $\Upsilon(u)=a$.
~\%player $u$ proposal is accepted}{\footnotesize \par}

\end{ALC@g}\STATE \textbf{\footnotesize{}else}{\footnotesize{} ~\%matched
to $u'\rightarrow\Upsilon(a)=u'$}{\footnotesize \par}

\begin{ALC@g}\STATE \textbf{\footnotesize{}if }{\footnotesize{}$u'\succ_{a}u$~\%player
$u$ proposal is rejected}{\footnotesize \par}

\begin{ALC@g}\STATE {\footnotesize{}player $u$ removes mmAP $a$
from its preference list}{\footnotesize \par}

\end{ALC@g}\STATE \textbf{\footnotesize{}else}{\footnotesize{}~\%player
$u$ proposal is accepted}{\footnotesize \par}

\begin{ALC@g}\STATE {\footnotesize{}$\Upsilon(a)=u$,$\Upsilon(u)=a$,
$\Upsilon(u')=\phi$}{\footnotesize \par}

\STATE {\footnotesize{}player $u'$ removes mmAP $a$ from its preference
list}{\footnotesize \par}

\end{ALC@g}\STATE \textbf{\footnotesize{}end if}{\footnotesize \par}

\end{ALC@g}\STATE \textbf{\footnotesize{}end if}{\footnotesize \par}

\STATE \textbf{\footnotesize{}if }{\footnotesize{}$\gamma_{u}<\gamma^{\textrm{Th}}$}{\footnotesize \par}

\begin{ALC@g}\STATE split $u$ into two players $u_{1}$ and $u_{2}$,
$\Upsilon(u_{2})=\phi$

\end{ALC@g}\STATE \textbf{\footnotesize{}end if}{\footnotesize \par}

\end{ALC@g}\STATE \textbf{\footnotesize{}until }{\footnotesize{}all
players are either matched with $\gamma_{u}\geq\gamma^{\textrm{Th}}$
or not having mmAPs that satisfy $a\succ_{u}u$ in their preference
lists}{\footnotesize \par}

\STATE {\footnotesize{} }\textbf{\footnotesize{}Output: }{\footnotesize{}a
stable matching $\Upsilon$}{\footnotesize \par}

\end{algorithmic}

\caption{\label{alg:matching} VRP-mmAP matching algorithm.}
\end{algorithm}

\begin{defn}
\label{def:matching}Given two disjoint sets of mmAPs and players
$(\mathcal{A},\mathcal{U}$), a \emph{matching} is defined as a \emph{one-to-one}
mapping $\Upsilon$ from the set $\mathcal{E}\cup\mathcal{U}$ into
the set of all subsets of $\mathcal{A}\cup\mathcal{U}$, such that
for each $a\in\mathcal{A}$ and $u\in\mathcal{U}$: \vspace{-0.1cm}
 
\end{defn}
\begin{enumerate}
\item $\forall u\in\mathcal{U},$$\Upsilon(u)\in\mathcal{A}\cup u$, where
$\Upsilon(u)=u$ means that a player is not associated to a remote
server, and will perform local LQ frame processing. 
\item $\forall a\in\mathcal{A},$$\Upsilon(a)\in\mathcal{U}\cup\{a\}$,
where $\Upsilon(a)=a$ means that the mmAP $a$ have no associated
players. 
\item $\mid\Upsilon(u)\mid=1,\mid\Upsilon(a)\mid=1$;~ 4)$\Upsilon(u)=a\Leftrightarrow\Upsilon(a)=u$. 
\end{enumerate}
By inspecting the problem in (\ref{eq:objective}), we can see that
the one-to-one mapping of the matching game satisfies the constraint
(\ref{eq:match_constraint_2}). Moreover, since preference profiles
can be defined to capture the cost function of the matching sets.
The utility of the mmAPs will essentially reflect the latency constraint
in (\ref{eq:final_delay_const}). Therefore, we define the utility
of associating player $u$ to mmAP $a$ as:

\vspace{-0.5cm}
 
\begin{equation}
\Phi_{au}(t)=D_{\textrm{th}}\epsilon-\sum_{i|V_{f}^{i}\in Q(t)}\frac{L_{fi}^{\textrm{HD}}}{T_{t}\bar{r}_{i}(t)}-\frac{\kappa L_{fu}^{\textrm{HD}}}{c_{e}}-\frac{L_{fu}^{\textrm{HD}}}{T_{t}\bar{r}_{u}(t)},\vspace{-0.2cm}
\end{equation}
\vspace{-0.1cm}
and the mmAP preference as: 
\begin{equation}
u\succ_{a}u'\Leftrightarrow\Phi_{au}(t)>\Phi_{au'}(t),\;a\succ_{a}u\Leftrightarrow\Phi_{au}(t)<0,\vspace{-0.1cm}\label{eq:cloudlet_preference}
\end{equation}
where the second preference states that a mmAP is not interested in
matching to a player that will violate its latency constraint. In
other words, the utility of each cloudlet is to seek a matching that
maximizes the difference between the right hand side and the left
hand side of the inequality in (\ref{eq:final_delay_const}), such
that the constraint is met as a stable matching is reached. To meet
the players' reliability target, we define the preference profiles
of the players as to maximize their link quality as follows: \vspace{-0.5cm}

\begin{equation}
\negthickspace a\negthinspace\succ_{u}\negthinspace a'\negthinspace\Leftrightarrow\negthinspace|h_{au}(t)|^{2}g_{au}^{\textup{Tx}}(t)g_{au}^{\textup{Rx}}(t)\negthinspace<\negthinspace|h_{a'u}(t)|^{2}g_{a'u}^{\textup{Tx}}(t)g_{a'u}^{\textup{Rx}}(t).\negthickspace\negthickspace\negthickspace\negthickspace
\label{eq:user_preference}
\end{equation}

Since users may not always find a single reliable link, we propose
to split the users that are matched but with an SINR below a predefined
threshold $\gamma^{\textrm{Th}}$ into multiple players, allowing
to be matched to multiple mmAPs and satisfy their link reliability
requirements subject to mmAP availability.

Next, matching stability is defined and an efficient multi-stage algorithm
based on deferred-acceptance (DA)\cite{gale_shapley} to solve it.
\begin{defn}
\label{def:blocking_pair}Given a matching $\Upsilon$ with $\Upsilon(a)=u$
and $\Upsilon(u)=a$, and a pair $(u',a')$ with $\Upsilon(a)\neq u'$
and $\Upsilon(u)\neq a'$, $(u',a')$ is said to be blocking the matching
$\Upsilon$ and form a blocking pair if: 1) $u'\succ_{a}u$, 2) $a'\succ_{u}a$.
A matching $\Upsilon*$ is stable if there is no blocking pair. {\setlength{\parindent}{0cm}\emph{ }

\emph{Remark} 1. The algorithm described in Algorithm \ref{alg:matching},
converges to a two-sided stable matching of players to mmAPs or to
their local servers \cite{gale_shapley}.} 
\end{defn}
\vspace{-0.1cm}

\section{Simulation Results\label{sec:SimRes}}

\vspace{-0.1cm}

In this section, we numerically validate the effectiveness of the
proposed solution. We also compare the proposed approach against two
benchmarking schemes: 
\begin{enumerate}
\item \emph{Baseline}~\emph{1}, with neither MC nor proactive computing
and caching. Requests are computed in real time after they are initiated,
and are transmitted through single-connectivity links. 
\item \emph{Baseline}~\emph{2}, without MC, but assumes proactive computing
and caching capabilities. 
\end{enumerate}
We consider a gaming arcade with a capacity of $8$x$8$ VR pods and
a set of default parameters\footnote{We consider $100$ impulse actions with popularity parameter $z=0.8$,
a uniformly distributed impact matrix $\boldsymbol{\theta}$, $D_{\textrm{th}}=100$
ms $\epsilon=0.1$, $10$ dBm mmAP transmit power, $L_{f}^{\textrm{HD}}$
= $\sim\exp(2)$ Gbit, $\kappa/c_{e}=5*10^{-8}$, $S/A=20$ video
frames, and $T_{\textrm{w}}=100$ ms.} unless stated otherwise. Impulse actions arrive following the Zipf
popularity model of parameter $z$ \cite{ejder_edge_2014}. Accordingly,
the arrival rate for the $i^{\textrm{th}}$ most popular action is
proportional to $1/i^{z}$.

\begin{figure*}
\hspace{0.3cm}%
\begin{minipage}[t]{0.4\paperwidth}%
\includegraphics[bb=10bp 0bp 390bp 295bp,clip,width=0.34\paperwidth]{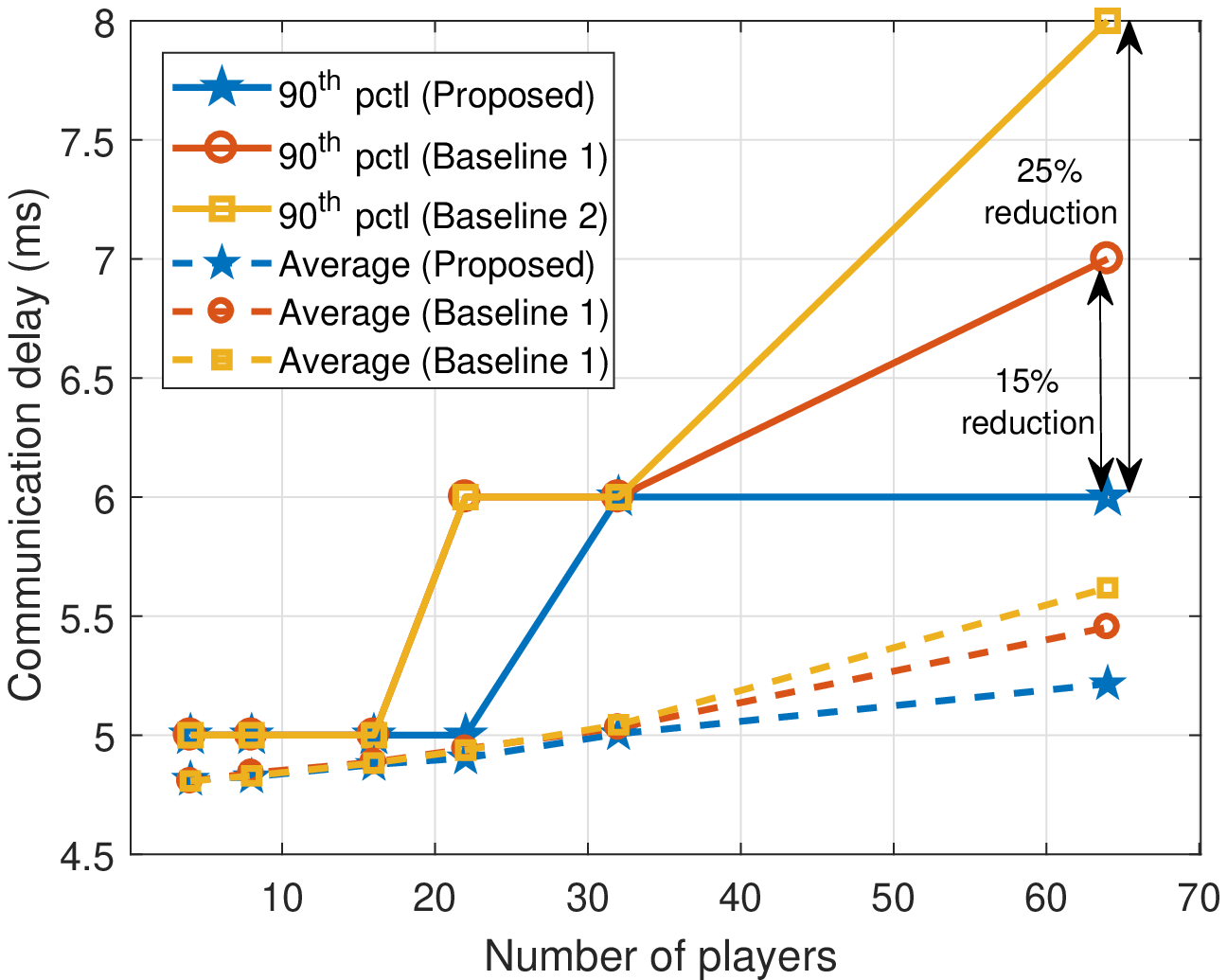}\caption{Average (dashed lines) and $90^{th}$ percentile (solid lines) communication
delay versus number of players, for $16$ mmAPs. \label{fig:90_delay_users}}
\end{minipage}\hspace{0.6cm}%
\begin{minipage}[t]{0.4\paperwidth}%
\includegraphics[bb=10bp 0bp 413bp 295bp,clip,width=0.36\paperwidth]{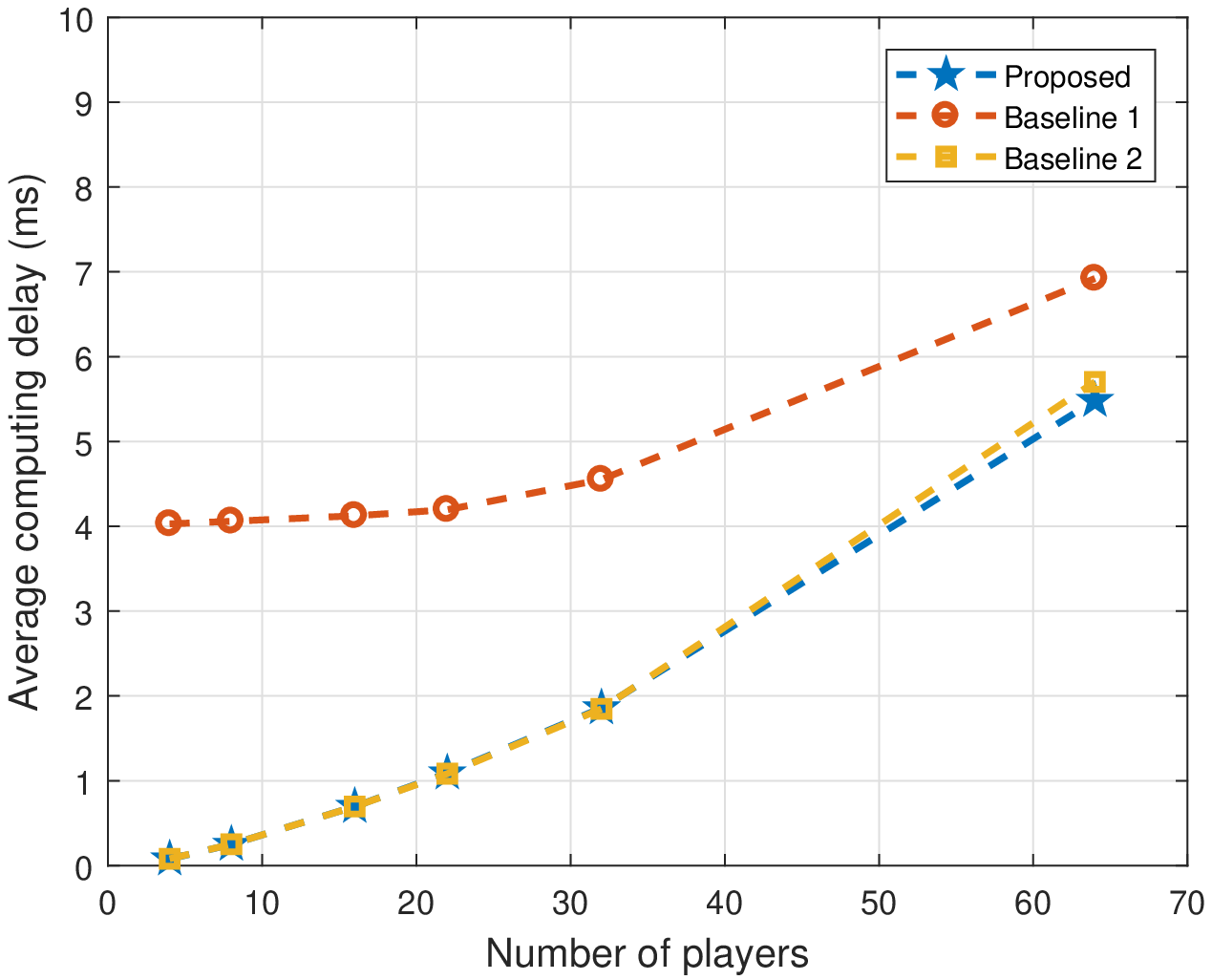}\caption{Average computing delay versus number of players, for $16$ mmAPs.\label{fig:two_delays_users}}
\end{minipage}
\end{figure*}

\begin{figure*}
\hspace{0.3cm}%
\begin{minipage}[t]{0.4\paperwidth}%
\hspace{0.3cm}\includegraphics[bb=20bp 0bp 400bp 295bp,clip,width=0.34\paperwidth]{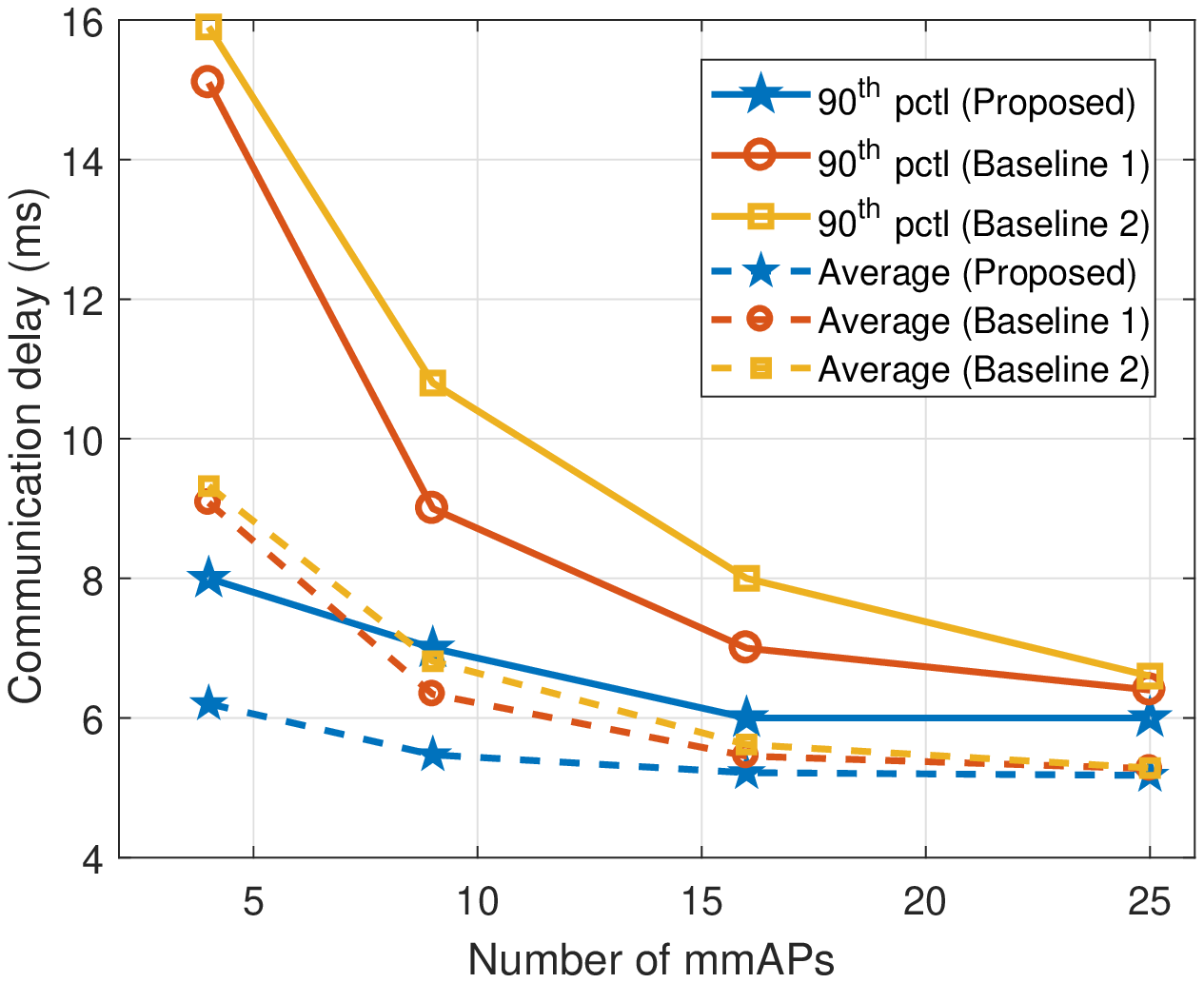}\caption{Average (dashed lines) and $90^{th}$ percentile (solid lines) communication
delay versus number of mmAPs, for $64$ players.\label{fig:90_delay_servers}}
\end{minipage}\hspace{0.5cm}%
\begin{minipage}[t]{0.4\paperwidth}%
\hspace{0.3cm}\includegraphics[bb=20bp 0bp 400bp 295bp,clip,width=0.35\paperwidth]{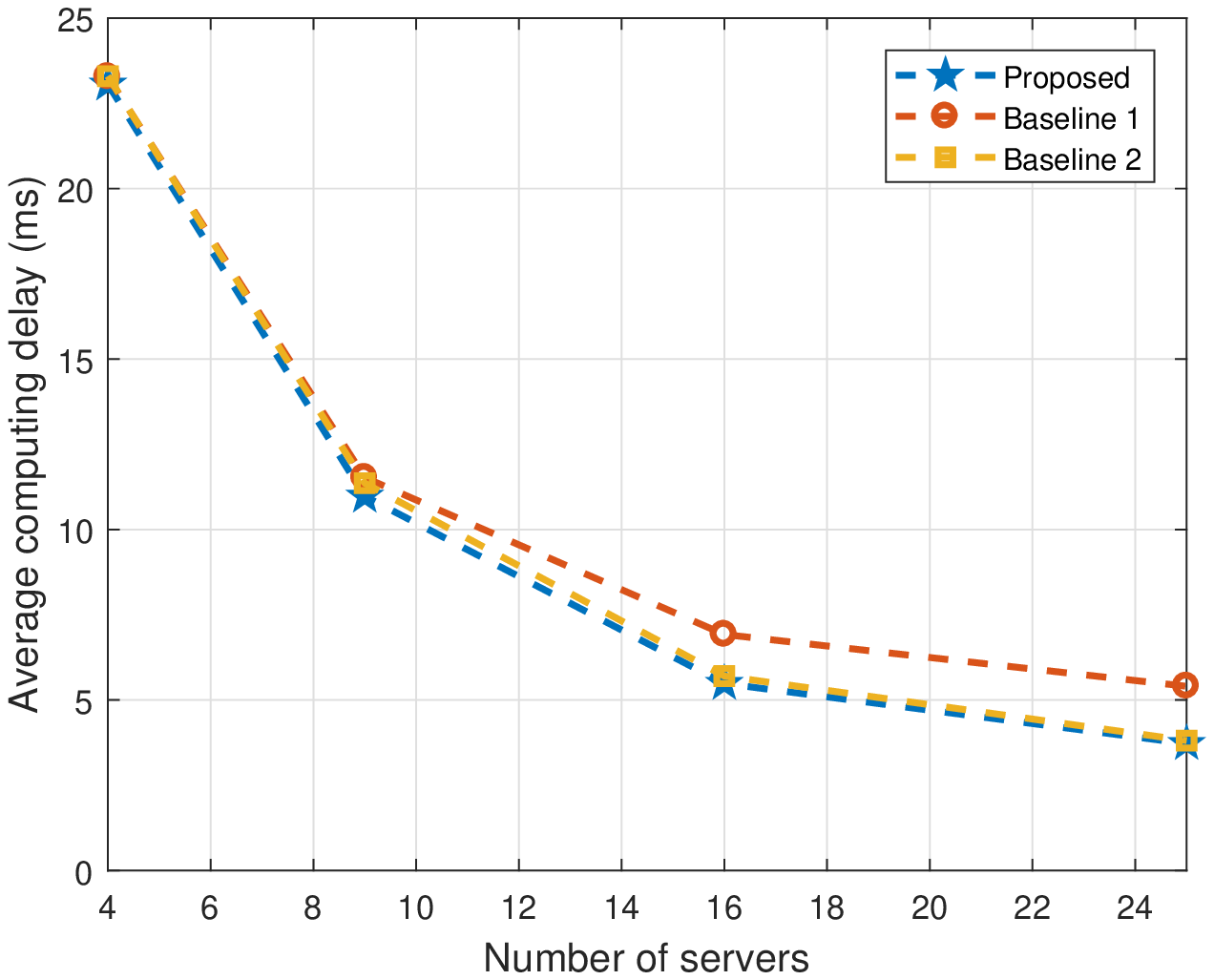}\caption{Average computing delay versus number of servers, for $64$ players.\label{fig:two_delays_servers}}
\end{minipage}
\end{figure*}


\subsection{Impact of number of players}

First, we investigate the performance of the proposed approach when
the number of players increases. We fix the number of mmAPs and servers
to $16$. In Fig.~\ref{fig:90_delay_users}, we show the average
and the $90^{th}$ percentile communication delay ($D_{uf}^{\textrm{cm}}$
90 pctl) for different schemes, which all increase with the network
density. This is due to the increase in offered load, as compared
to the network capacity, and the higher levels of interference. As
the number of players increases, the proposed approach achieves up
to $25\%$ reduction in the $D_{uf}^{\textrm{cm}}$ 90 pctl, due to
the MC gain that allows users with weak links to receive from multiple
servers. This reduction is more evident in dense network conditions
associated with high interference/blockage levels. In Fig.~\ref{fig:two_delays_users},
we show the change in the average computing delay. Both the proposed
approach and Baseline 2 achieve significant reduction in the computing
delay, due to leveraging the proactivity to cut down the rendering
latency. 


\subsection{Impact of number of mmAPs and servers}


Next, we investigate the performance of the proposed approach as the
number of mmAPs and servers increase, while fixing the number of players
to the maximum arcade capacity of $64$. We assume the number of servers
in the fog network also matches the number of mmAPs. First, Fig.~\ref{fig:90_delay_servers}
shows the average and $D_{uf}^{\textrm{cm}}$ 90 pctl communication
delay performance. Intuitively, low number of mmAPs will incur higher
communication delay due to having higher offered load than what the
mmAPs can serve. However, we observe that at low number of mmAPs,
the average delay can be reduced by up to $33\%$ whereas the $D_{uf}^{\textrm{cm}}$
90 pctl can be halved using the proposed approach.

Finally, Fig.~\ref{fig:two_delays_servers}, shows the computing
and communication delay performance. We can see that, at low number
of servers, the computing delay is always high, due to not having
enough computing resources to serve the high number of players. The
network can hardly serve the real-time computing requests, leaving
no room for proactive computing. Accordingly, higher number of servers
are needed to achieve proactive computing gains. 

\vspace{-0.1cm}

\section{Conclusions}

\label{sec:Conc} 

In this paper, we have studied the problem of ultra-reliable and low
latency wireless VR networks. A joint proactive computing and user
association scheme is proposed in mmWave enabled VR for interactive
gaming. In the proposed scheme, information about the game players'
upcoming pose and game action is leveraged to proactively render their
HD video frames such that computing latency is minimized. To ensure
reliable and low latency communication, a matching algorithm has been
proposed to associate players to mmAPs and enable multi-connectivity,
in which multiple mmAPs jointly transmit the video frames to players
to overcome the effect of channel variability. Simulation results
have shown that the proposed scheme achieves significant reduction
in both computing and communication latency under different network
conditions, as compared to different baseline schemes. \vspace{-0.2cm}

\section{Acknowledgments}

\label{sec:ack} 

This research was partially supported by the Academy of Finland project
CARMA, the NOKIA donation project FOGGY and by the Spanish MINECO
under grant TEC2016-80090-C2-2-R (5RANVIR). \vspace{-0.5cm}
\bibliographystyle{IEEEtran}


\end{document}